\documentclass[twoside,english,iop]{emulateapj}
\usepackage[T1]{fontenc}
\usepackage{amssymb}
\usepackage{graphicx}

\makeatletter
\usepackage{apjfonts}
\usepackage[hyperfootnotes=false,colorlinks=true,citecolor=blue]{hyperref}

\usepackage{apjfonts}
\usepackage[hyperfootnotes=false,colorlinks=true,citecolor=blue]{hyperref}
\shorttitle{DUST SCATTERING IN TURBULENT MEDIA}
\shortauthors{SEON \& WITT}

\makeatother

\usepackage{babel}
\begin{document}

\title{DUST SCATTERING IN TURBULENT MEDIA:\\ CORRELATION BETWEEN THE SCATTERED
LIGHT AND DUST COLUMN DENSITY}

\author{Kwang-Il Seon\altaffilmark{1,2} and Adolf N. Witt\altaffilmark{3}}

\altaffiltext{1}{Korea Astronomy and Space Science Institute, Daejeon 305-348, Republic of Korea; kiseon@kasi.re.kr}
\altaffiltext{2}{Astronomy and Space Science Major, University of Science and Technology, Daejeon 305-350, Republic of Korea}
\altaffiltext{2}{Ritter Astrophysical Research Center, University of Toledo, Toledo, OH 43606, USA} 
\begin{abstract}
Radiative transfer models in a spherical, turbulent interstellar medium
(ISM) in which the photon source is situated at the center are calculated
to investigate the correlation between the scattered light and the
dust column density. The medium is modeled using fractional Brownian
motion structures that are appropriate for turbulent ISM. The correlation
plot between the scattered light and optical depth shows substantial
scatter and deviation from simple proportionality. It was also found
that the overall density contrast is smoothed out in scattered light.
In other words, there is an enhancement of the dust-scattered flux
in low-density regions, while the scattered flux is suppressed in
high-density regions. The correlation becomes less significant as
the scattering becomes closer to be isotropic and the medium becomes
more turbulent. Therefore, the scattered light observed in near-infrared
wavelengths would show much weaker correlation than the observations
in optical and ultraviolet wavelengths. We also find that the correlation
plot between scattered lights at two different wavelengths shows a
tighter correlation than that of the scattered light versus the optical
depth.
\end{abstract}

\keywords{dust, extinction --- methods: numerical --- radiative transfer ---
scattering}

\section{INTRODUCTION}

It is known that dust-scattered light is proportional to the dust
column density in the optically thin regions. The correlation between
the dust-scattered light and tracers of the dust column density such
as far-infrared (FIR) emission and neutral hydrogen column density,
has been extensively examined in near-infrared (NIR), optical and
far-ultraviolet (FUV) wavelengths \citep{Guhathakurta1989,Bowyer91,Schiminovich2001,Witt2008,Seon2011a,Brandt2012,Malinen2013}.
However, it was found that the correlation relation shows large scatter
\citep[e.g.,][]{Ienaka2013}. The correlation slope between the scattered
light and the dust tracers was also found to vary from sightline to
sightline. Various causes (including spatial variations of the illuminating
radiation field, the dust temperature, and the dust grain properties
and/or the contribution from emission lines) have been suggested as
factors that can explain the large variation in the correlation relationship.
Recently, it was found that a substantial fraction of the scatter
is attributable to the lognormal probability density function (PDF)
of the density in a turbulent interstellar medium (ISM) \citep{Seon2011a,Seon2013}.

The ISM is turbulent and the PDF of structure is hierarchical, scale-free,
and/or fractal \citep{Elmegreen2004}. The PDFs of the densities of
the turbulent ISM have been found to be close to lognormal not only
in numerical simulations \citep{Vazquez94,Klessen2000,Ostriker01,Burkhart2012}
but also in observations \citep{Padoan97,Berkhuijsen2008,Lombardi2008,Seon2009,Froebrich2010}.

Radiative transfer can significantly differ in homogeneous and clumpy
media because photons can easily escape through low-density regions
\citep{Witt1996,Witt2000,Mathis2002}. It is crucial, therefore, to
use a realistic density structure of the dusty ISM in the radiative
transfer models. Most previous radiative transfer models in the dusty
ISM have assumed uniform or smoothly varying density distributions
\citep{Witt1992,Bianchi1996,DeLooze_Somberero2012}, or a two-phase
medium to represent a clumpy medium \citep{Witt1996,Witt2000,Bianchi2000,Stalevski2012}.
\citet{Witt1997} adopted a spectrum of three types of clouds to describe
the scattering medium (see also \citealt{Schiminovich2001}). A more
complex algorithm to mimic hierarchically clumped clouds proposed
by \citet{Elmegreen1997} has been used for radiative transfer models
in the dusty ISM \citep{Mathis2002}. However, this hierarchical model
does not satisfy the condition of a lognormal PDF.

The purpose of this paper is to mimic the ISM density structure such
that it matches as closely as possible the current best knowledge
and to investigate the general properties of dust radiative transfers.
We use the fractional Brownian motion (fBm) algorithm to simulate
the turbulent ISM structure and perform Monte Carlo radiative transfer
calculations. This paper is organized as follows. In Section 2, we
review the density structure and radiative transfer code. Section
3 presents the effect of the turbulent density structure on dust scattering.
A summary is given in Section 4.

\section{MODEL}

A lognormal density field in the turbulent ISM can be characterized
by the standard deviation ($\sigma_{\ln\rho}$) or variance ($\sigma_{\ln\rho}^{2}$)
of the logarithm of the density and the power-law index ($\gamma$)
of the density power spectrum \citep{Seon2012}. In hydrodynamic regimes,
the variance of the log-density has been found to be related to the
sonic Mach number $M_{{\rm s}}$ of the medium according to 
\begin{equation}
\sigma_{\ln\rho}^{2}=\ln(1+b^{2}M_{{\rm s}}^{2}),\label{eq:1}
\end{equation}
where the proportional constant $b$, depending on the type of turbulence
forcing mode, is $1/3$, 0.4, or 1.0 for solenoidal, natural mixing,
and compressive modes, respectively, \citep{Federrath2008,Federrath2010}.
\citet{Seon2009} combines the numerical simulation results of \citet{Padoan04},
\citet{Kim05}, and \citet{Kritsuk06}, finding a relationship between
$M_{{\rm s}}$ and $\gamma$,
\begin{equation}
\gamma=-3.81M_{{\rm s}}^{-0.16},\label{eq:2}
\end{equation}
which may be applicable to solenoidal and natural mixing modes (see
also \citealt{Seon2012}). Therefore, given a Mach number, we can
determine the standard deviation and power spectral index.

A lognormal density field $\rho(\mathbf{x})$ is obtained by extending
the method in \citet{Elmegreen2002}. First, a Gaussian random field
$\rho_{{\rm g}}$($\mathbf{x}$) is generated using the fBm algorithm
\citep{Voss1988,Stutzki1998}. The fBm structures are generated by
assigning three-dimensional Fourier coefficients. The Fourier phases
are generated such that they are uniformly distributed and the amplitudes
are distributed in a Gaussian form with the variance $\mathbf{|k|}^{-\gamma_{{\rm g}}}$.
The inverse Fourier transform gives a Gaussian random field $\rho_{{\rm g}}$($\mathbf{x}$).
We then multiply the density field with the desired standard deviation
$\sigma_{\ln\rho}$ of the logarithmic density and exponentiate the
field to obtain $\rho(\mathbf{x})=\exp(\sigma_{\ln\rho}\rho_{{\rm g}})$.
Here, we should note that the power-law spectral index $\gamma$ of
the resulting lognormal density field is different from the spectral
index $\gamma_{{\rm g}}$ of the input Gaussian field. \citet{Seon2012}
derived the input spectral index $\gamma_{{\rm g}}$ as a polynomial
function of $(\gamma,\sigma_{\ln\rho})$. We use this relationship
together with Equations (\ref{eq:1}) and (\ref{eq:2}) to obtain
a random realization of the lognormal density field with the spectral
index $\gamma$ and standard deviation $\sigma_{\ln\rho}$ appropriate
for a given Mach number.

We assumed the natural mixing mode with $b=0.4$ in which solenoidal
and compressive forces are mixed naturally. A random realization of
the lognormal density field with box size of $N^{3}=256^{3}$ was
generated for each Mach number of $M_{{\rm s}}$ = 1.0 and 3.0.

For the dust-scattering simulation, we used the three-dimensional
radiative transfer code MoCafe (\textbf{Mo}nte \textbf{Ca}rlo radiative
trans\textbf{fe}r) \citep{Seon2009,Jo2012,SeonWitt2012,Lim2013,Choi2013}.
The direction into which a photon is scattered is randomly determined
from a Henyey-Greenstein phase function, based on the algorithm of
\citet{Witt1977I}. The first scattering of the photons is forced
to ensure that every photon contributes to the scattered light \citep{Cashwell1959}.
Photons are assumed to originate at a centrally located, isotropically
radiating point source and to escape from the dust cloud at a spherical
surface with a diameter equal to the box size. The spherical surface
was divided into equal areas using the HEALPix (Hierarchical Equal
Area isoLatitude Pixelization) scheme with a resolution parameter
$N_{{\rm side}}=16$, corresponding to an angular resolution of $\sim3.7^{\circ}$.

The homogeneous optical depth $\tau_{{\rm H}}$ for a lognormal density
cloud is defined by the optical depth of a cloud with a constant density,
but with the same dust mass as the lognormal density cloud within
the sphere. For each dust cloud, $\tau_{{\rm H}}$ was varied from
0.1 to 1.4 and the asymmetry factor $g$ ranged from 0.0 to 0.99.
We obtained the fluxes of scattered light ($F_{{\rm scatt}}$) measured
at every pixel on the sphere and the optical depths ($\tau$) integrated
from the center to the spherical surface. We also calculated a series
of uniform density models with the optical depth ranging from 0.1
to 1.4. The results for lognormal density clouds were then compared
with those of the uniform density clouds. An albedo of $a=0.5$ is
assumed throughout the paper unless noted otherwise. When other albedos
were used, we obtained similar results except that the overall levels
of scattered light were scaled up or down according to the albedo.
We used $10^{8}$ photons in each of the models.

A fractal ISM structure is scale-free and contains regions which are
much denser than the mean density. Thus, it is essential to use a
large number of photons to probe the complex structure through the
scattering. A grid with $N^{3}=256^{3}$ may not be appropriate to
resolve the density gradients in small scales. To examine these effects,
we used $10^{10}$ photons for a few cases. However, no differences
larger than 2\% were found. We also generated lognormal density fields
with a box size of $N^{3}=512{}^{3}$, and binned the data cube into
coarser box sizes. The radiative transfer results obtained with box
sizes of down to $128^{3}$ were essentially the same as those of
$512^{3}$. However, noticeable differences were found in the models
with $N^{3}\le32{}^{3}$. Therefore, our simulations are well suited
for the radiative transfer calculations in the complex density structures.

\section{RESULTS}

\begin{figure*}[t]
\begin{centering}
\includegraphics[angle=90,scale=0.15]{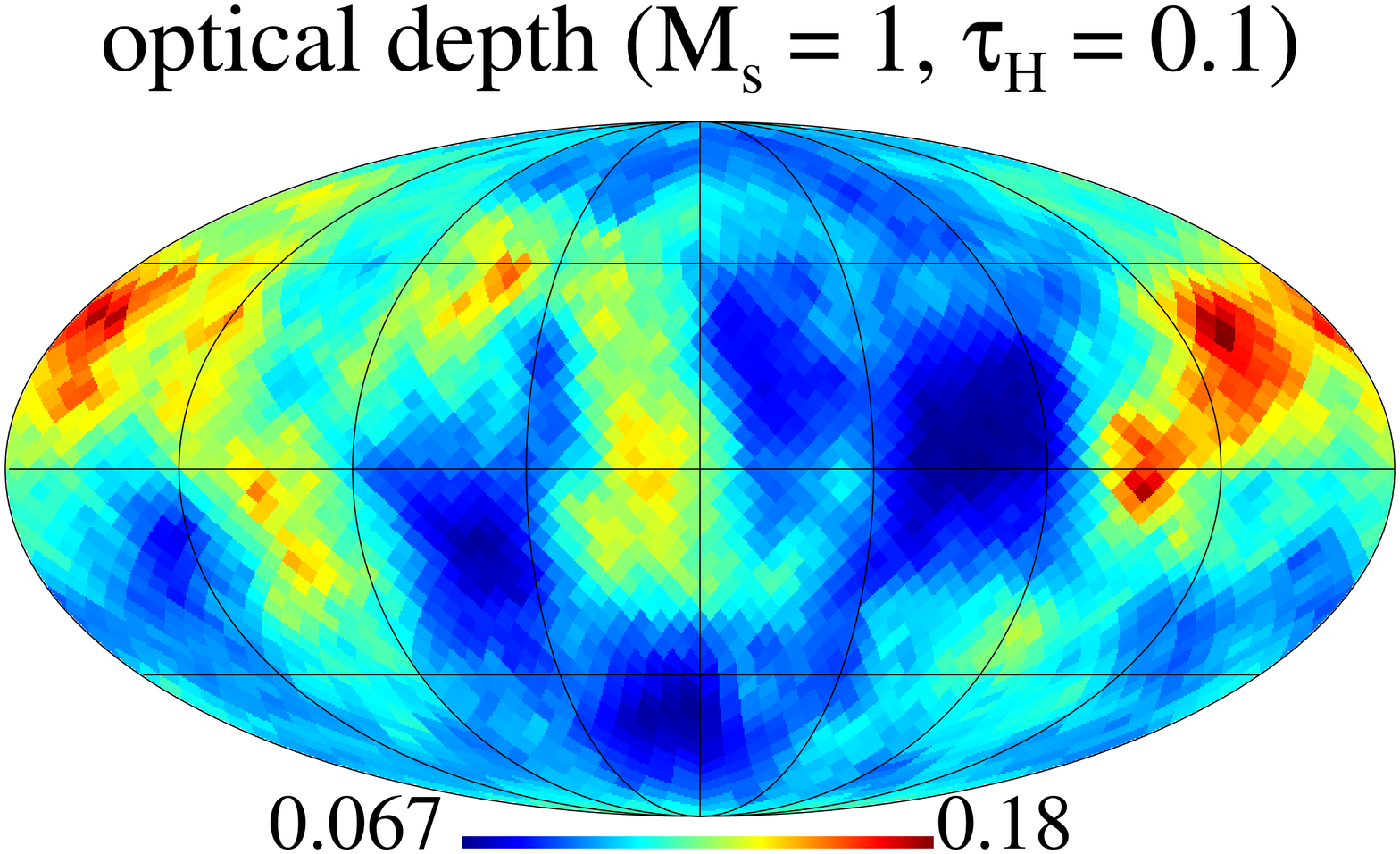} \includegraphics[scale=0.15,angle=90]{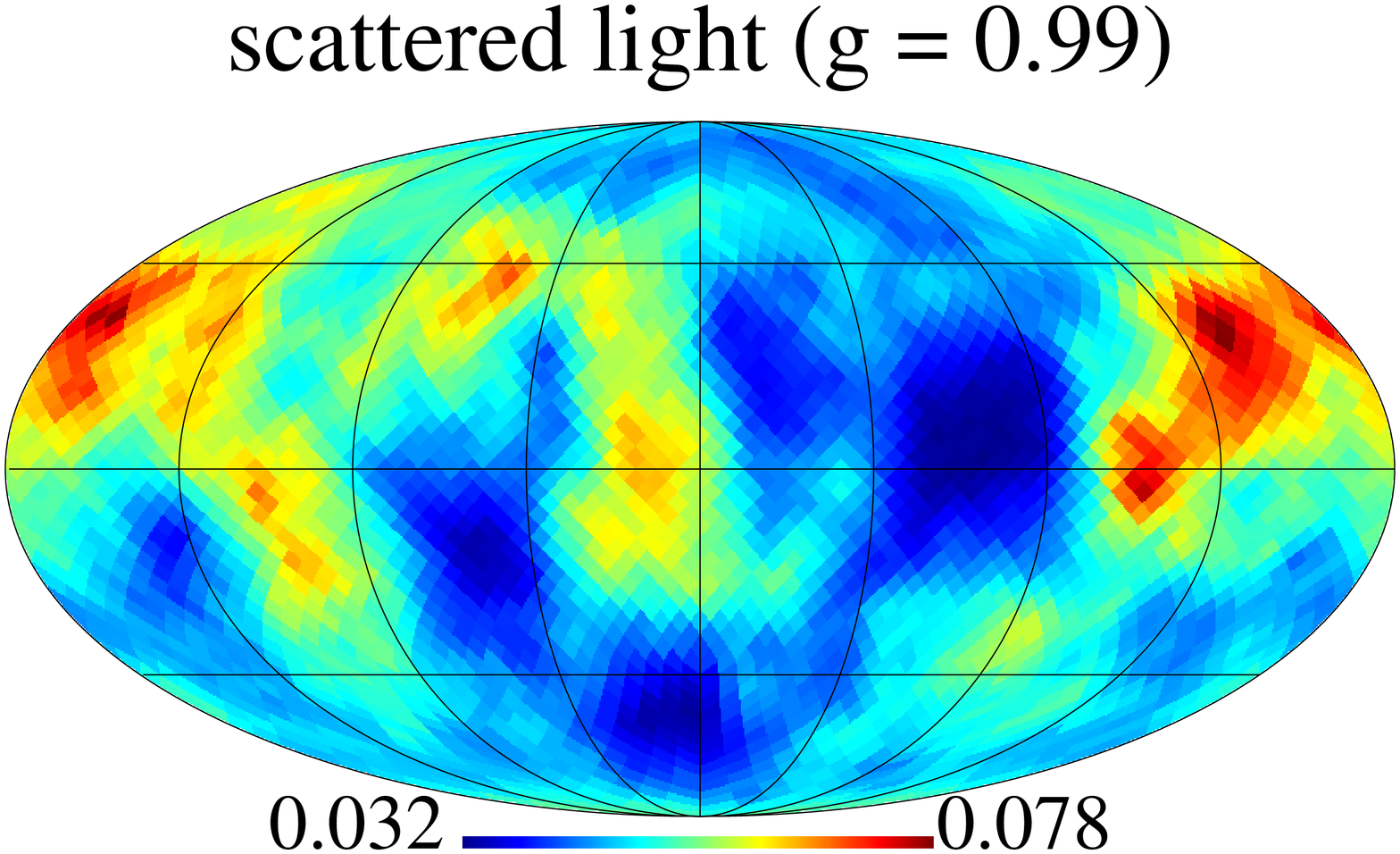}
\includegraphics[scale=0.15,angle=90]{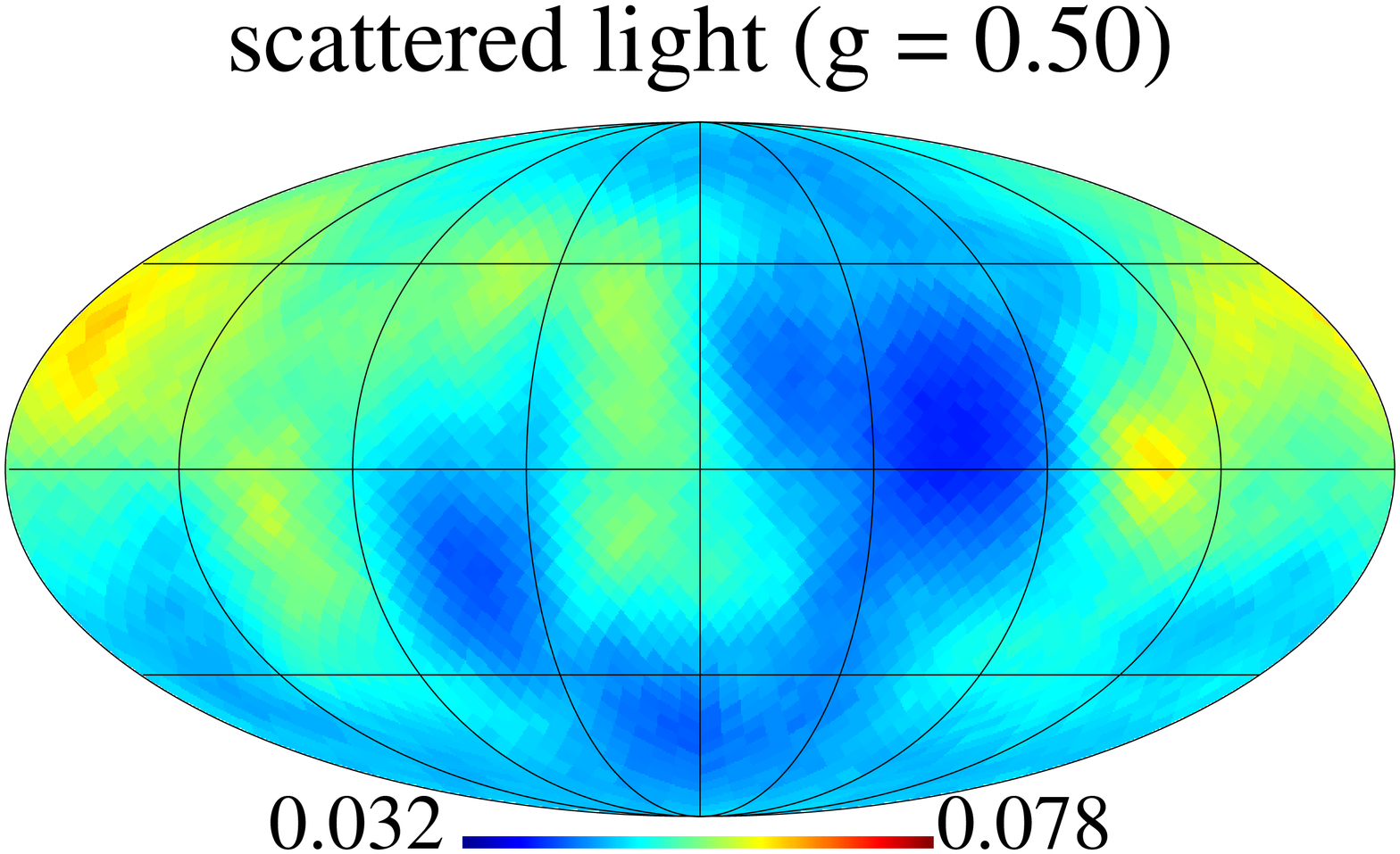} \includegraphics[scale=0.15,angle=90]{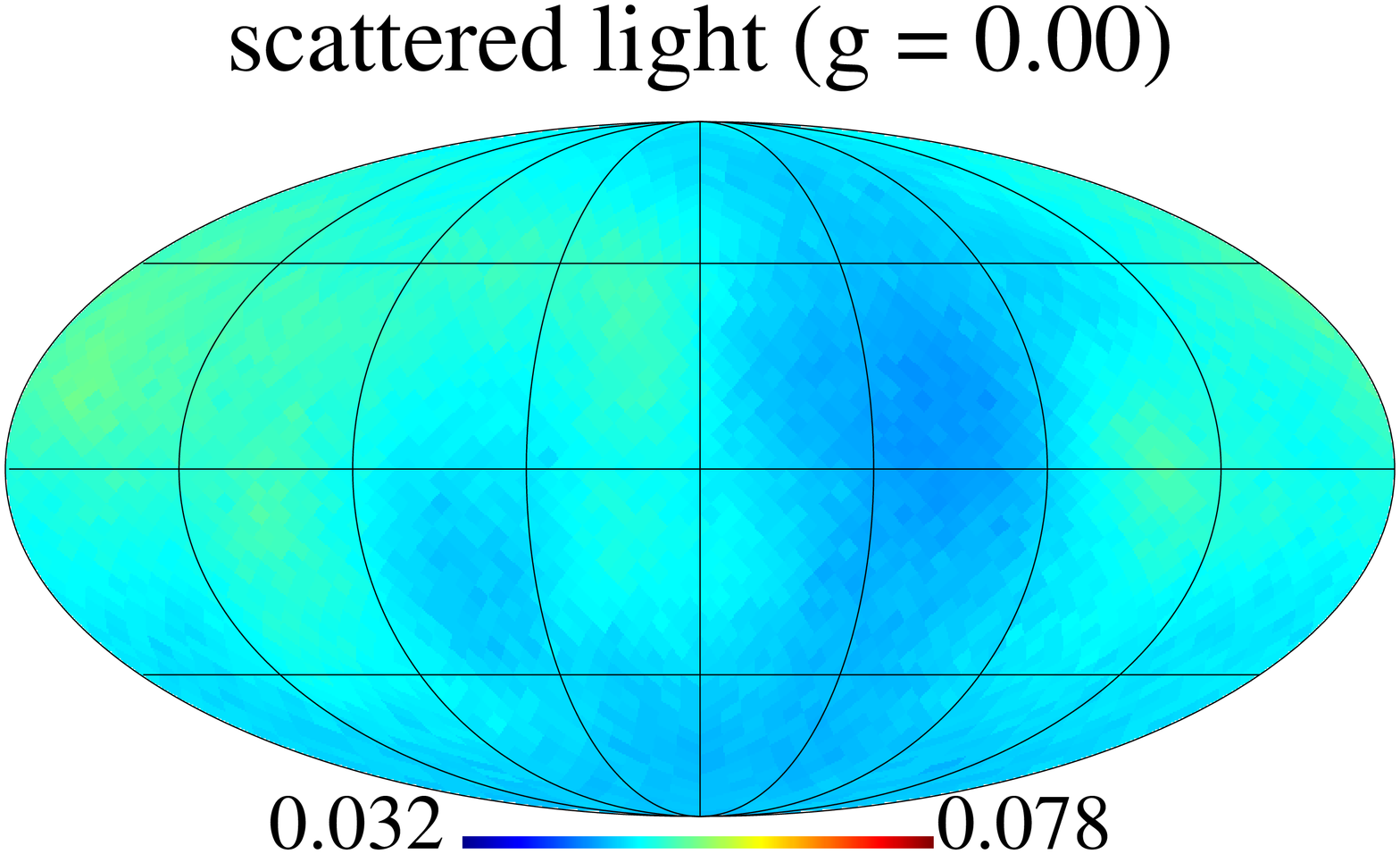}
\par\end{centering}

\begin{centering}
\medskip{}

\par\end{centering}

\begin{centering}
\includegraphics[angle=90,scale=0.15]{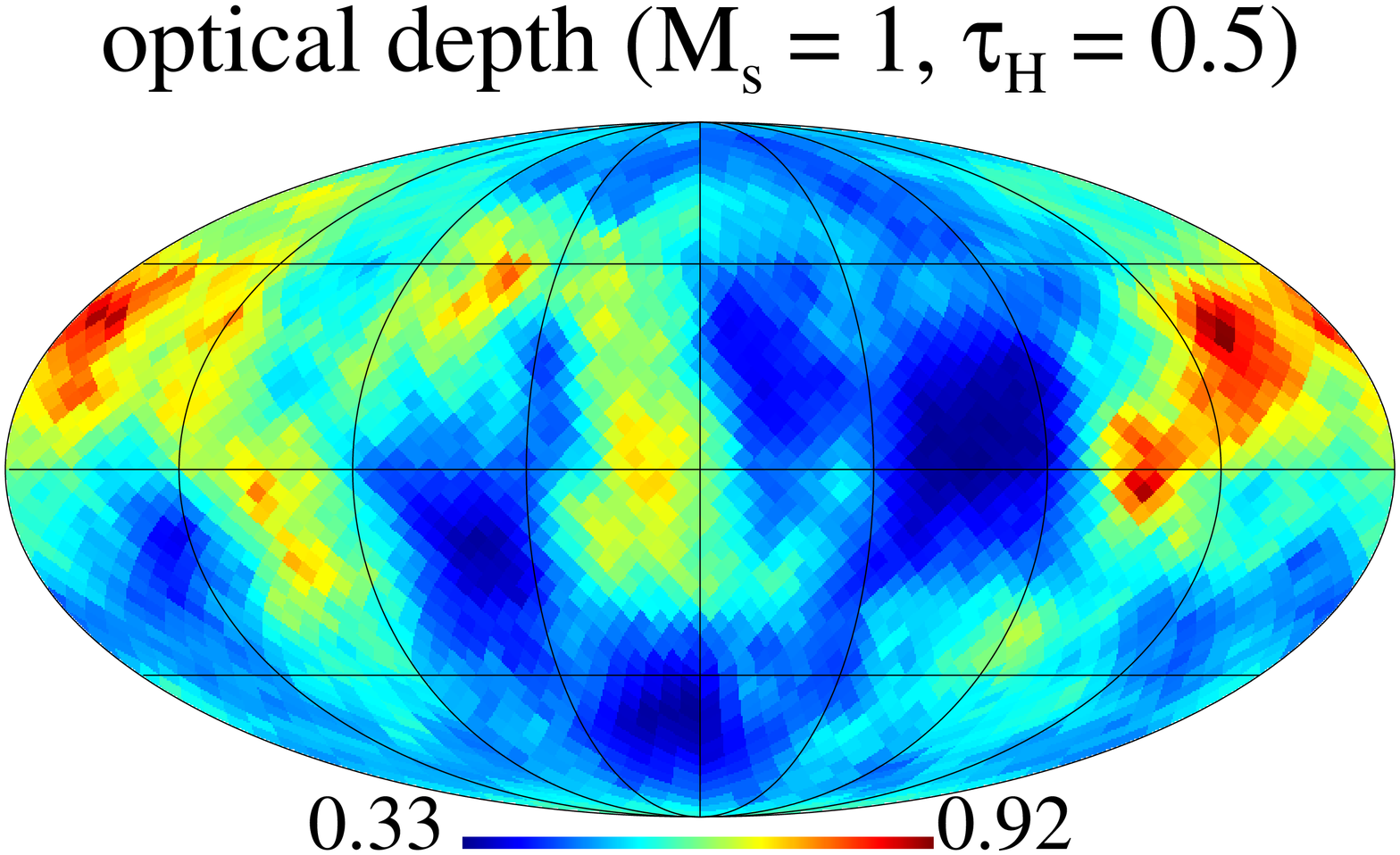} \includegraphics[scale=0.15,angle=90]{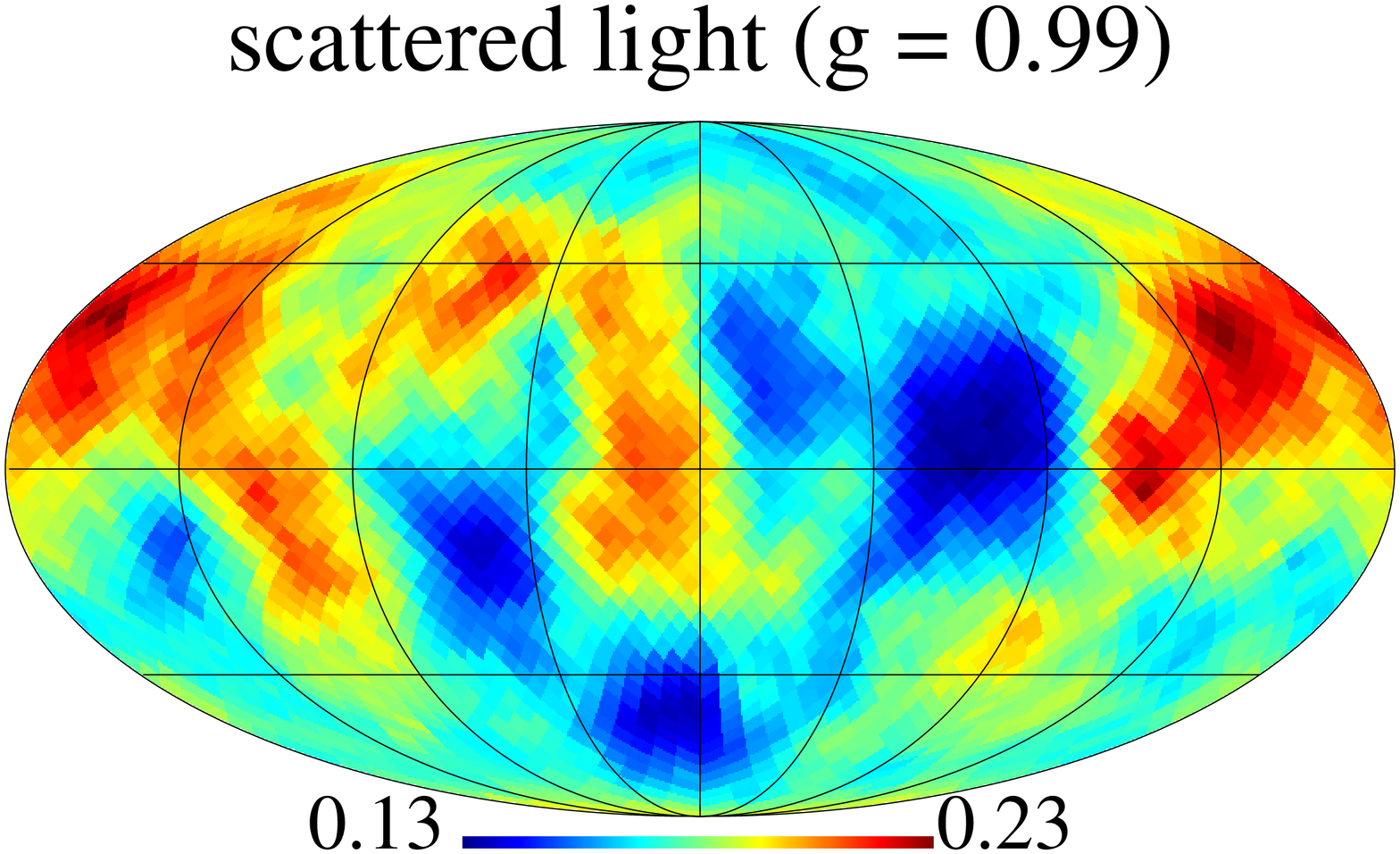}
\includegraphics[scale=0.15,angle=90]{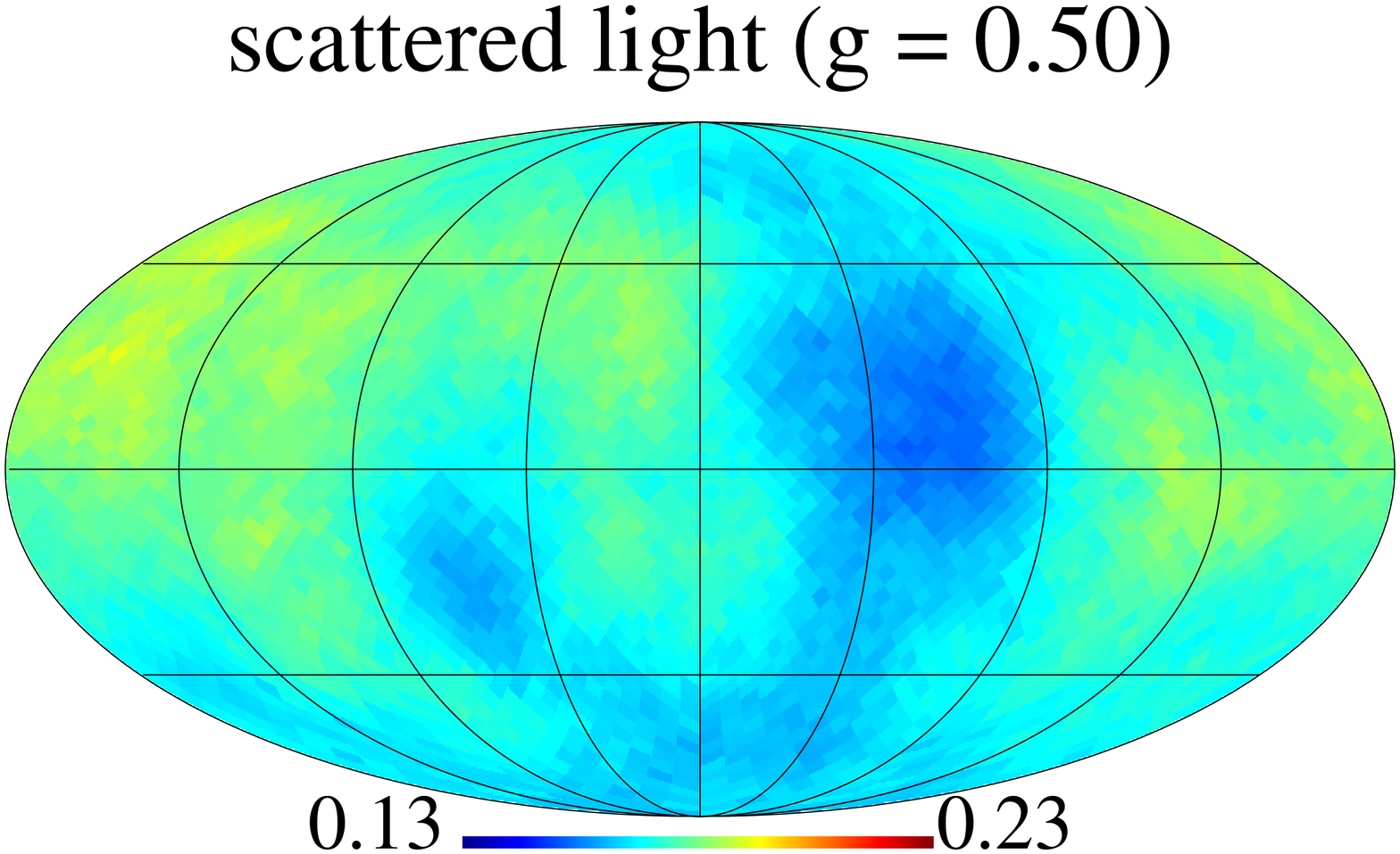} \includegraphics[scale=0.15,angle=90]{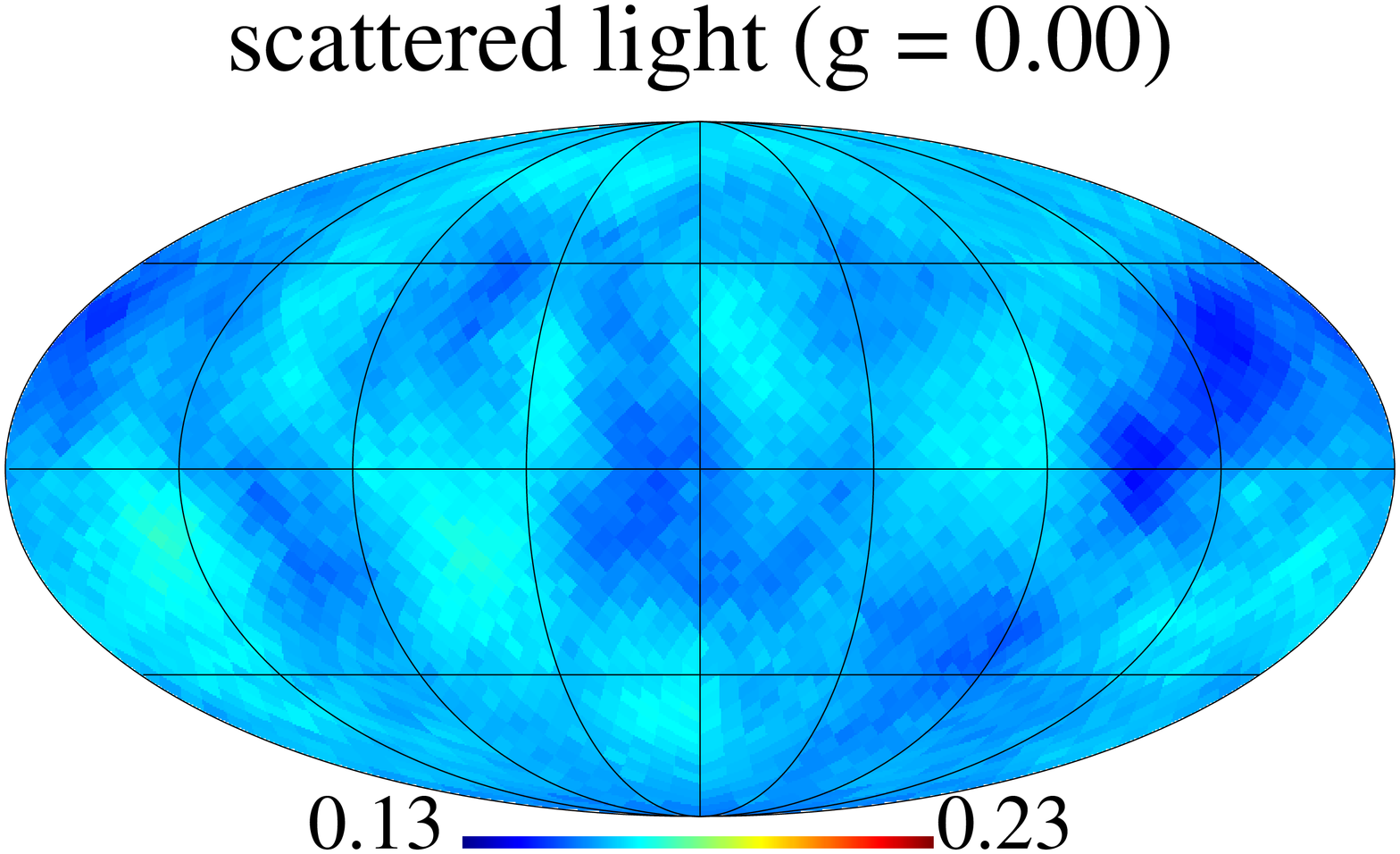}
\par\end{centering}

\begin{centering}
\medskip{}

\par\end{centering}

\caption{\label{images-M1}Mollweide maps of the optical depths and scattered
fluxes obtained for the Mach number of $M_{{\rm s}}=1$. Maps in the
first to fourth rows show the results for the homogenous optical depths
of $\tau_{{\rm H}}$ = 0.1 and 0.5, respectively. For each $\tau_{{\rm H}}$,
the first column shows maps of the optical depth, and the second,
third, and fourth columns present maps of the scattered light obtained
when the asymmetry phase factor ($g$) is 0.99, 0.5, or 0.0, respectively.
For each $\tau_{{\rm H}}$, the same color contrasts of the scattered
fluxes are used for comparison.}
\end{figure*}

\begin{figure*}[t]
\begin{centering}
\includegraphics[angle=90,scale=0.15]{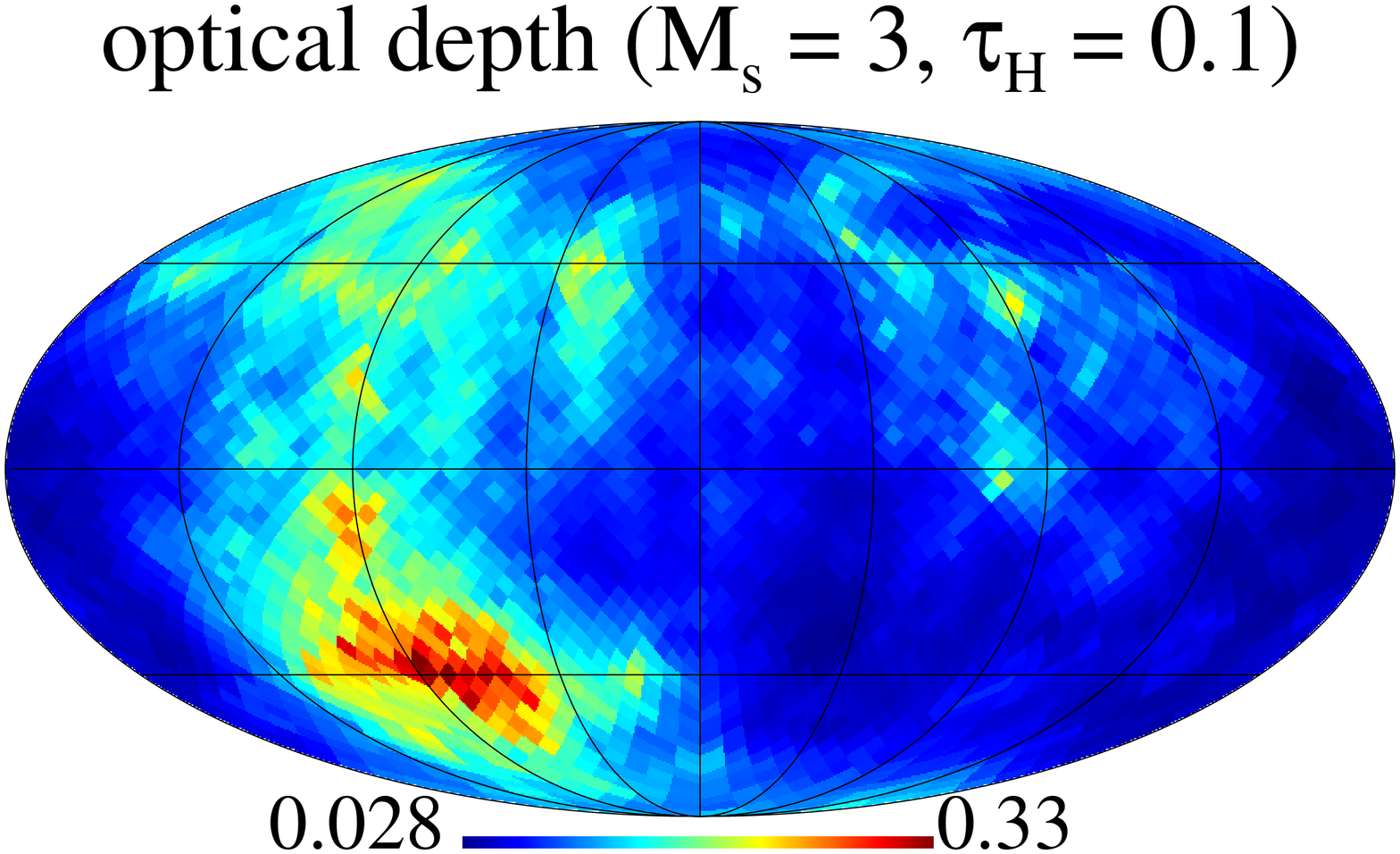} \includegraphics[scale=0.15,angle=90]{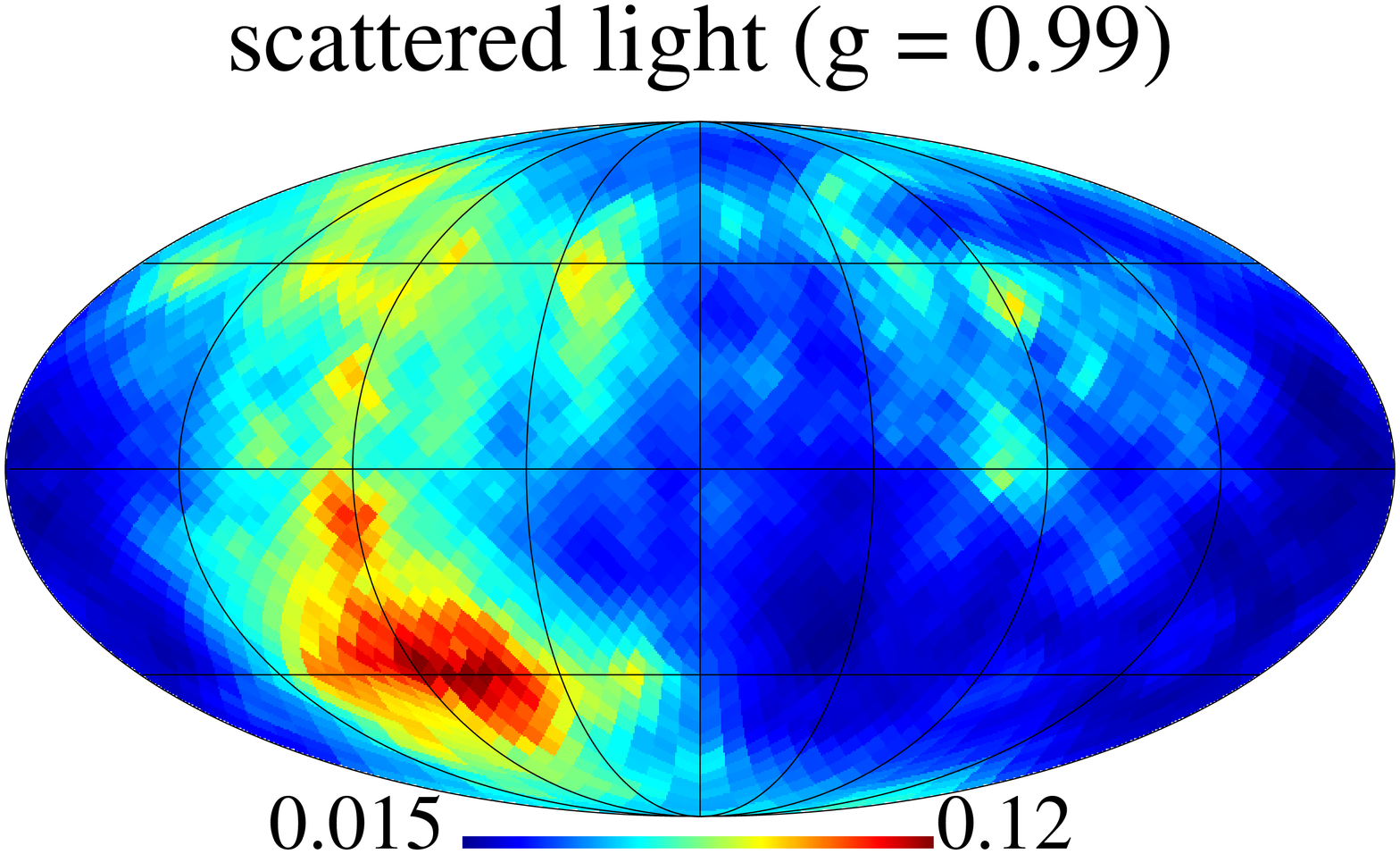}
\includegraphics[scale=0.15,angle=90]{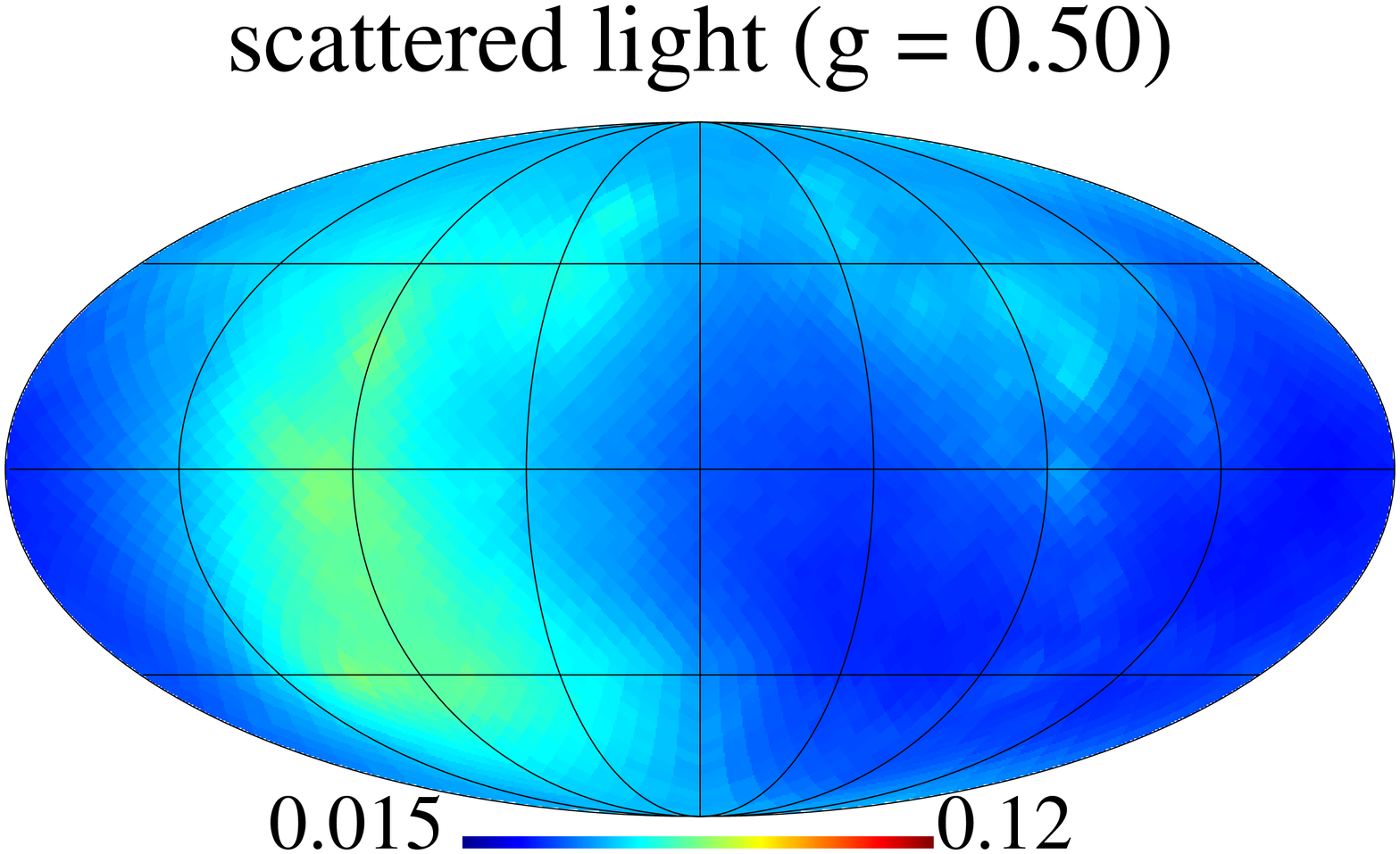} \includegraphics[scale=0.15,angle=90]{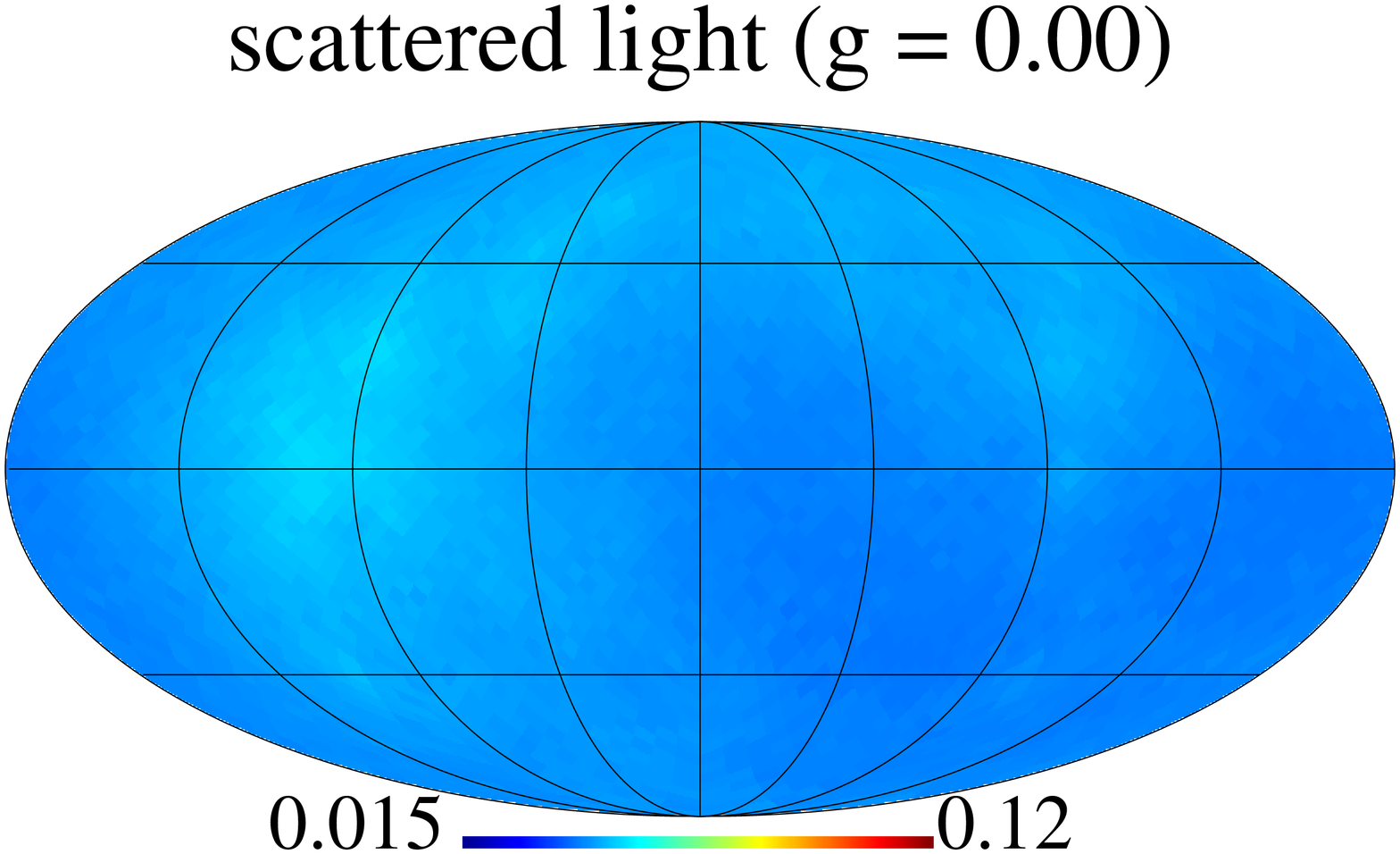}
\par\end{centering}

\begin{centering}
\medskip{}

\par\end{centering}

\begin{centering}
\includegraphics[angle=90,scale=0.15]{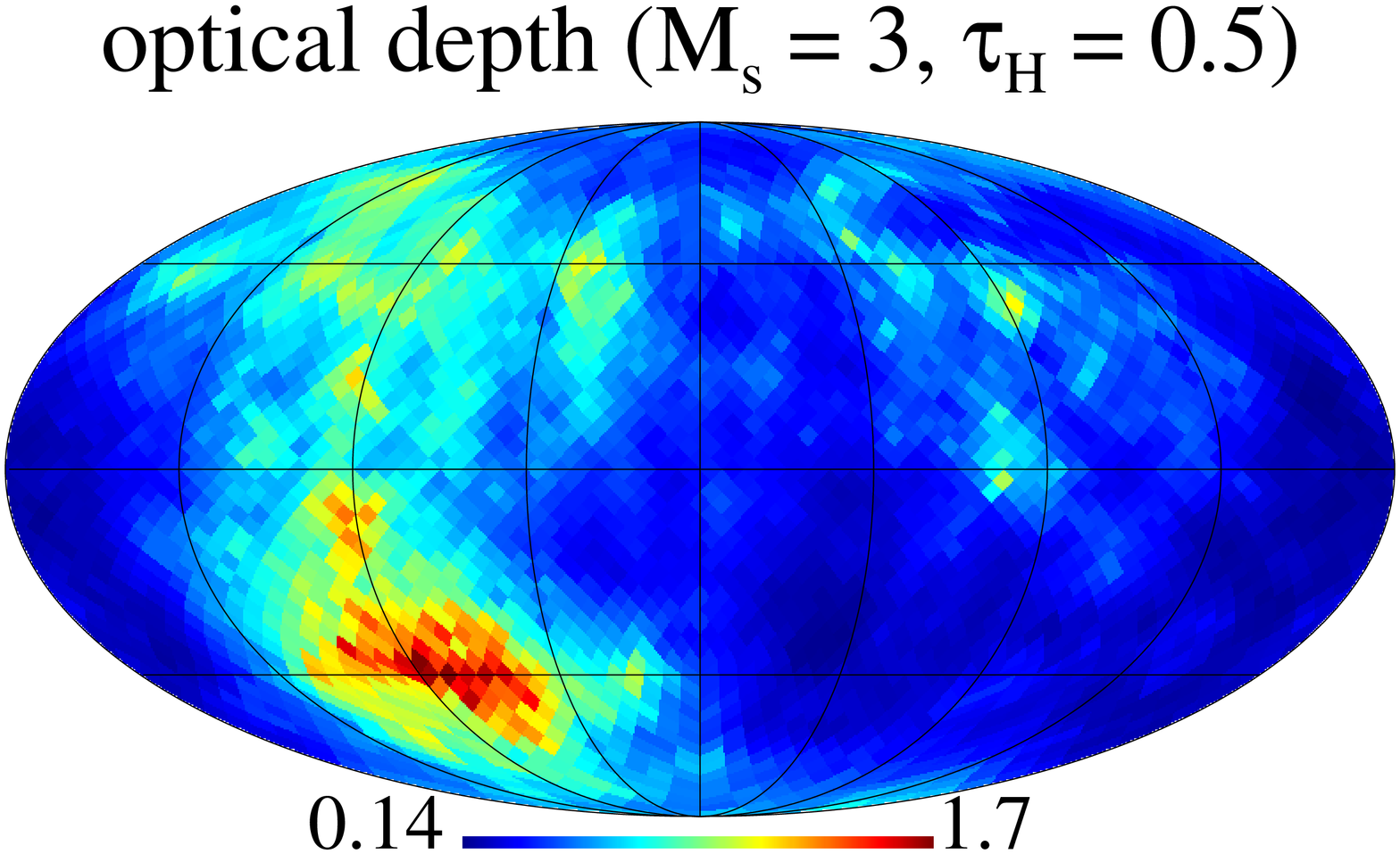} \includegraphics[scale=0.15,angle=90]{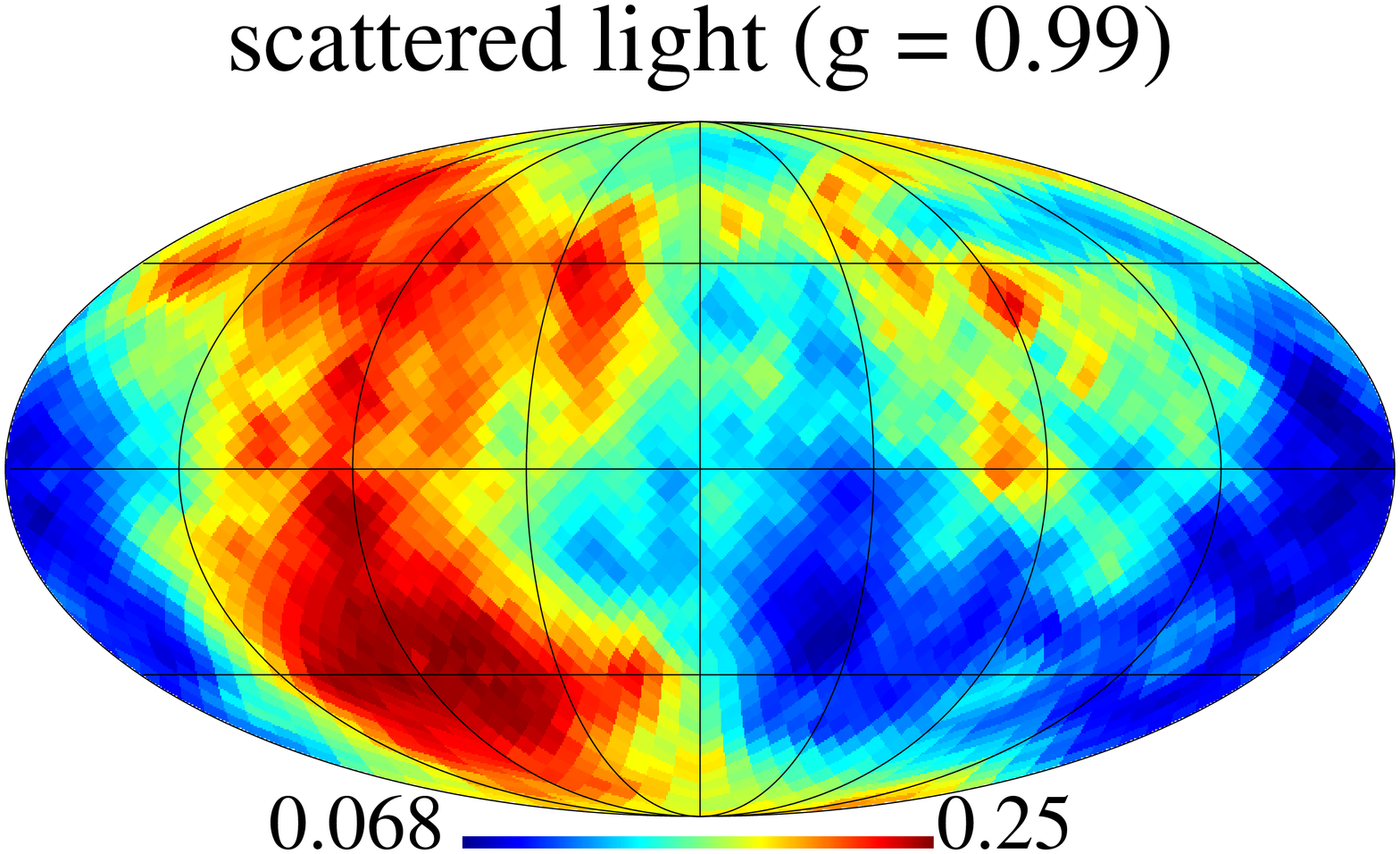}
\includegraphics[scale=0.15,angle=90]{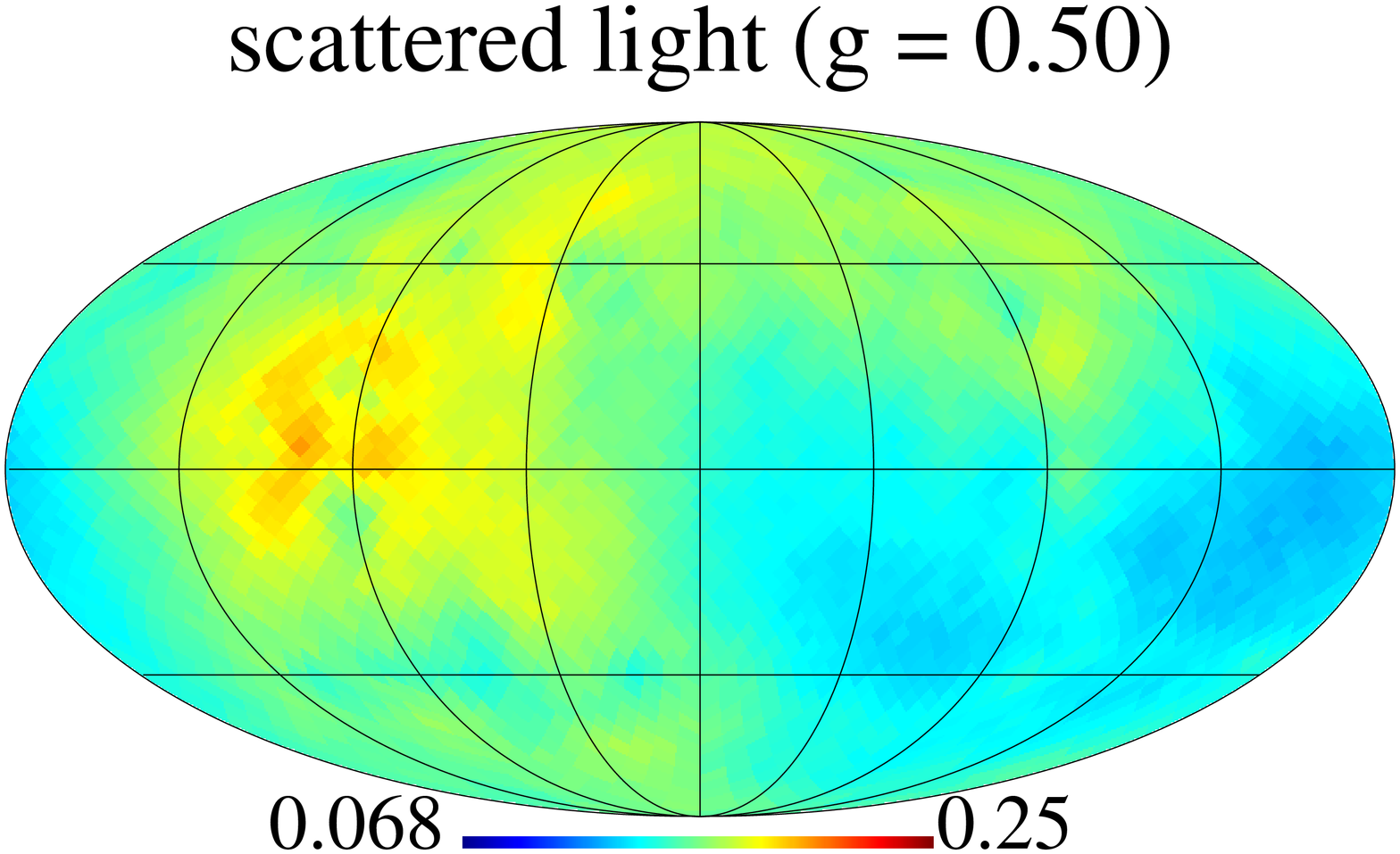} \includegraphics[scale=0.15,angle=90]{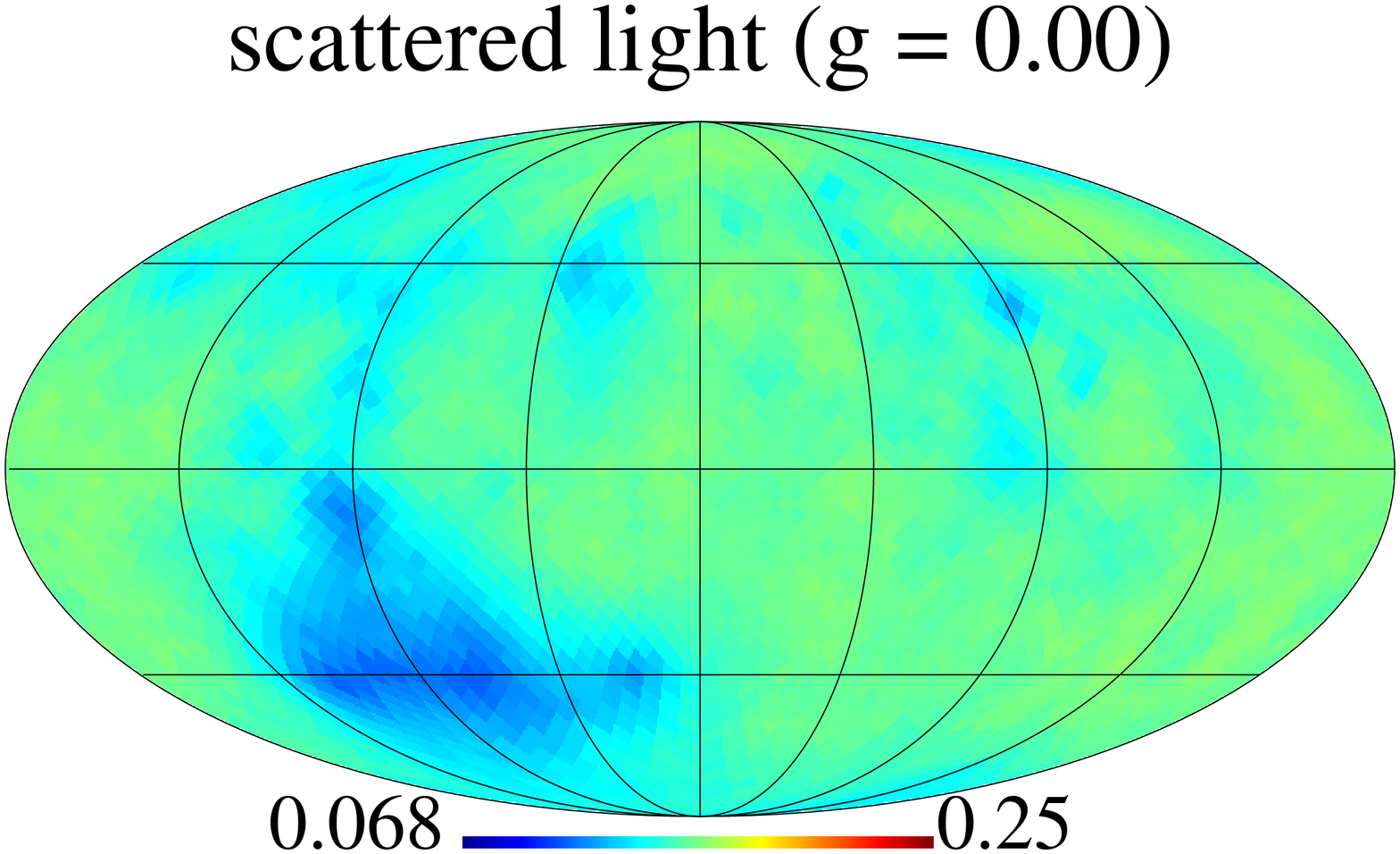}
\par\end{centering}

\begin{centering}
\medskip{}

\par\end{centering}

\caption{\label{images-M3}Mollweide maps of the optical depths and scattered
fluxes obtained for the Mach number of $M_{{\rm s}}=3$.}
\end{figure*}

Figures \ref{images-M1} and \ref{images-M3} show the Mollweide projection
maps of $\tau$ and $F_{{\rm scatt}}$ for the clouds with $\tau_{{\rm H}}$
values of 0.1 and 0.5 for each of $M_{{\rm s}}$ values of 1 and 3.
For each $(M_{{\rm s}},\;\tau_{{\rm H}})$ combination, the radiative
transfer models with three asymmetry factors ($g=0.99$, 0.5, and
0.0) are presented to emphasize the differences in the correlations.
The three asymmetry factors represent almost-complete forward scattering
(0.99), typical forward scattering (0.5) in optical and UV wavelengths,
and isotropic scattering (0.0), which may be applicable to NIR wavelengths
longer than $\sim2\mu$m. The scattered flux is expressed relative
to the flux $F_{0}$ that would be observed in the absence of dust.

As shown in Figures \ref{images-M1} and \ref{images-M3}, the correlation
weakens gradually as $g$ decreases and $M_{{\rm s}}$ increases.
The most dramatic change in the correlation is found when $g$ decreases.
Therefore, the results are discussed in the order of a decreasing
value of $g$. The scattered light for $g=0.99$ correlates very well
with $\tau$, although a slightly weaker correlation is found for
large values of $\tau_{{\rm H}}$ and $M_{{\rm s}}$. When $g=0.5$,
the contrast of scattered light is largely smoothed out. When $(M_{{\rm s}},\;\tau_{{\rm H}},\; g)=(3,\;0.5,\;0.5)$,
weak anti-correlations become apparent at locally dense sightlines.
When $g=0$, the correlation between $F_{{\rm scatt}}$ and $\tau$
is scarcely seen, with even anti-correlations found. When $(\tau_{{\rm H}},\; g)=(0.5,\;0)$,
the positive correlations completely disappear. Instead, weak anti-correlations
are clearly shown. An anti-correlation at locally dense sightlines
with $\tau>1$, as shown when $(M_{{\rm s}},\;\tau_{{\rm H}},\; g)=(3,\;0.5,\;0)$,
$ $is not unexpected. However, the anti-correlation at sightlines
with $\tau<1$, shown when $(M_{{\rm s}},\;\tau_{{\rm H}},\; g)=(1,\;0.5,\;0)$,
cannot be understood generally in that the scattered light is correlated
with the optical depth at a low optical depth.

\begin{figure}[t]
\begin{centering}
\includegraphics[scale=0.68]{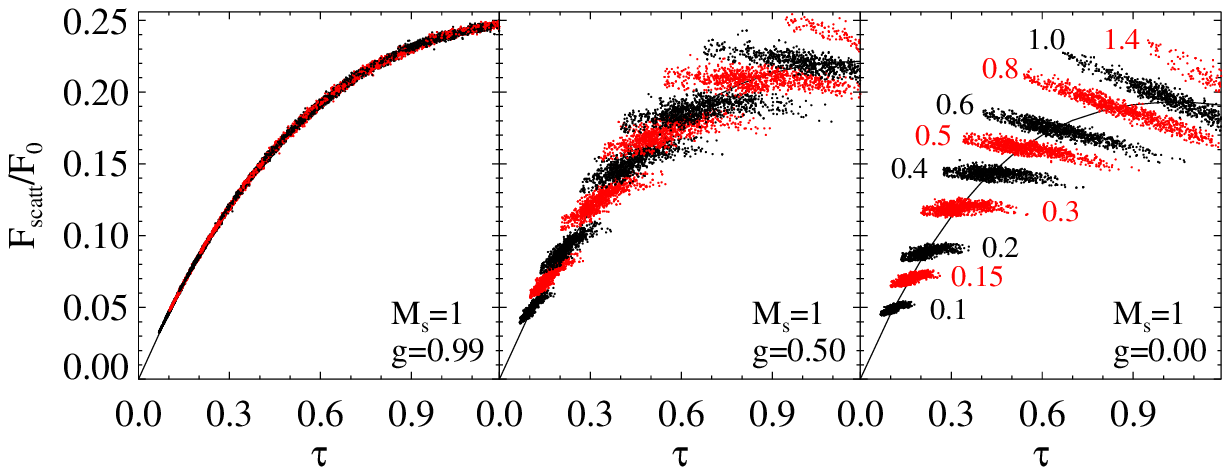} 
\par\end{centering}

\begin{centering}
\medskip{}

\par\end{centering}

\begin{centering}
\includegraphics[scale=0.68]{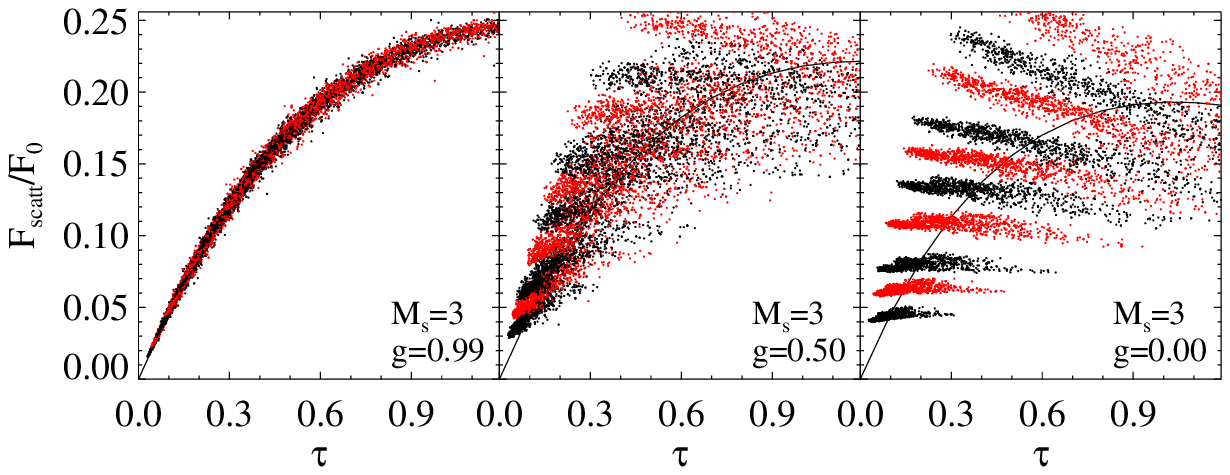} 
\par\end{centering}

\caption{\label{correlation}Scattered light versus the optical depth. For
each $(M_{{\rm s}},\; g)$, $\tau_{{\rm H}}$ varied from 0.1 to 1.4,
as denoted by the alternating colors in the plot of $(M_{{\rm s}},\; g)$
= (1, 0.0). Solid curves are the correlation plots of the uniform
density clouds. }
\end{figure}

These trends are shown more clearly in Figure \ref{correlation}.
For each of the six combinations of $(M_{{\rm s}},\; g)$, $\tau_{{\rm H}}$
is varied from 0.1 to 1.4 and the resulting correlations are shown
in alternating colors. The scattered fluxes, as functions of $\tau$,
from uniform density clouds are also shown in solid curves (these
are not clearly visible when $g=0.99$). First, we note large scatter
in the correlation plots, which becomes more evident when a medium
with a higher value of $M_{{\rm s}}$ is considered. Second, the scattered
flux in the sightlines with a low $\tau$ ($\lesssim\tau_{{\rm H}}$)
is enhanced compared to uniform density models while the flux in the
sightlines with a high value of $\tau$ ($\gtrsim\tau_{{\rm H}}$)
is suppressed. In other words, the correlation slope of each model
is shallower than the curves of uniform density models or has negative
value. This effect becomes more significant as $M_{{\rm s}}$ increases
and $g$ decreases. Third, an anti-correlation is found at a lower
$\tau_{{\rm H}}$ than that expected in uniform density models \citep{Witt1982}.
The critical $\tau_{{\rm H}}$ at which the anti-correlation occurs
is lower for a higher $M_{{\rm s}}$ and a lower $g$. For models
with $(M_{{\rm s}},\; g)=(3,\;0)$, for instance, no obvious positive
correlation is seen, even in the model with the lowest $\tau_{{\rm H}}$
($=0.1$). When $g=0$, the anti-correlation appears even in the models
with $\tau_{{\rm H}}=0.4$ despite the fact that the values of the
local optical depth $\tau$ are typically less than 1.

The above results can be understood as follows. In the case of complete
forward scattering ($g=1$), the scattered light will continue to
propagate along the original incident direction. Therefore, the scattered
intensity depends solely on the optical depth in the line of sight.
However, as $g$ decreases, the chance for photons to be scattered
in other directions increases. Regions with relatively high densities
would scatter photons into nearby regions of lower density. Low-density
regions would also scatter photons into higher density regions in
the vicinity, but this occurs less efficiently than the case of scattering
from high-density regions into low-density regions because the scattering
probability at a point is proportional to the density at that point.
Therefore, the final effect of non-complete forward scattering would
be a net flow of radiation from relatively dense regions to less dense
regions. At the optical thin limit, this effect results in a weaker
correlation (with a shallower slope) than uniform density clouds.
Anti-correlation is also found at a lower $\tau_{{\rm H}}$ compared
to the uniform density models. In the optically thick cases, this
gives rise to a stronger (steeper) anti-correlation. Clearly, the
effect becomes more significant as $g$ decreases. Moreover, the effect
will be more prominent in a medium with high density contrast, i.e.,
with a higher $M_{{\rm s}}$.

\begin{figure}[t]
\begin{centering}
\includegraphics[scale=0.68]{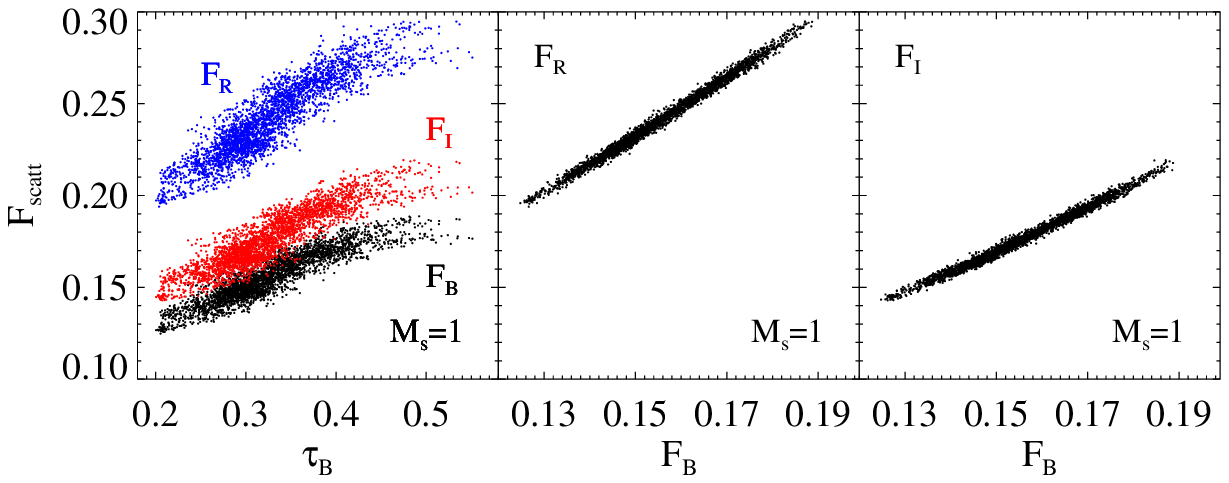} 
\par\end{centering}

\begin{centering}
\medskip{}

\par\end{centering}

\begin{centering}
\includegraphics[scale=0.68]{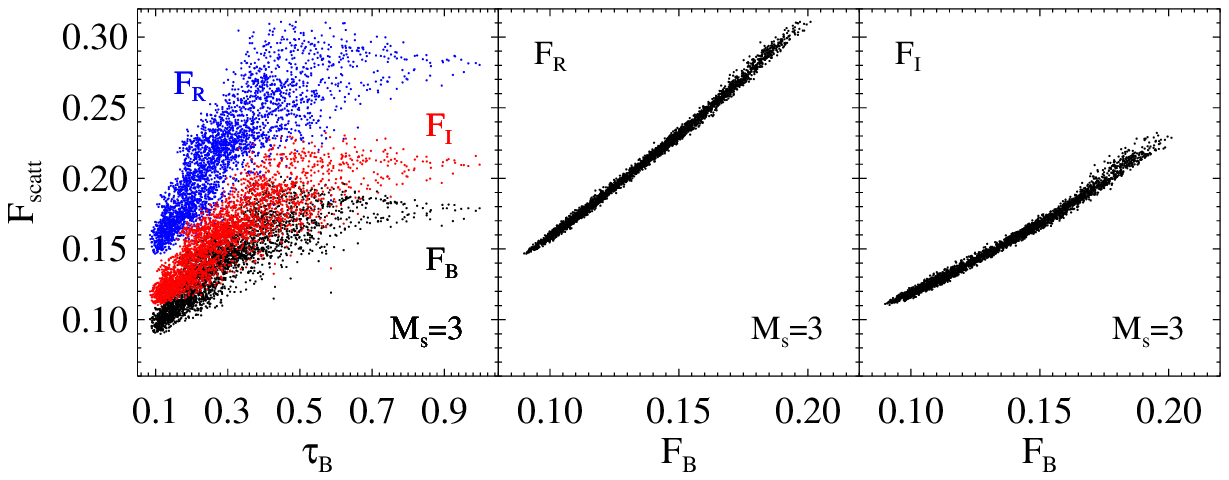}
\par\end{centering}

\caption{\label{BRI}Correlations between the scattered fluxes at three wavelengths,
roughly corresponding to the $B_{{\rm J}}$, $R$, and $I$ bands,
and the optical depth at the $B_{{\rm J}}$ band. We assumed $(\tau_{{\rm H}},\; g,\ a,\; F_{0})$
= (0.3, 0.6, 0.6, 1.02), (0.2, 0.5, 0.6, 2.19), and (0.1, 0.4, 0.5,
3.56) for the $B_{{\rm J}}$, $R$, and $I$ bands, respectively.
Here, $F_{0}$ is the input radiation field strength of \citet{MMP1983}.}
\end{figure}

We now compare our results with optical observations of cirrus clouds.
\citet{Guhathakurta1989} found that the correlation plots between
the $B_{{\rm J}}$ and $R$ bands intensities versus the 100 $\mu$m
intensity show shallower correlation slopes compared to predictions
obtained with a constant density model. However, the correlation slope
between the $I$ band and the 100 $\mu$m intensity showed a steeper
slope than the theoretical value. This may due to the large contribution
of the extended red emission in the $I$ band. The $B$ band correlation
plot of the four clouds observed in \citet{Witt2008} shows a slightly
steeper slope than the clouds of \citet{Guhathakurta1989}, but it
is still shallower than their constant density model. The $B$ band
correlation slope of the cloud observed in \citet{Ienaka2013} is
also shallower than the constant density model. \citet{Guhathakurta1989}
also found that the correlation between the scattered intensities
at two different bands show a tighter correlation than those between
the scattered light and the optical depth. To compare this with our
models, we examined a cloud with $\tau_{{\rm H}}=0.3$ at $B_{{\rm J}}$,
assuming that $a=$ (0.6, 0.6, 0.5), $g=$ (0.6, 0.5, 0.4), and $\tau_{{\rm H}}=$
(0.3, 0.2, 0.1) in the $B_{{\rm J}}$, $R$, and $I$ bands, respectively,
which are appropriate for the Milky Way dust in \citet{Draine03}.
We also assumed the radiation field strengths $F_{0}=$ (1.02, 2.19,
3.56) of \citet{MMP1983} at the three bands, as in \citet{Guhathakurta1989}.
Figure \ref{BRI} shows that the correlations between the fluxes at
the two bands are much stronger than the relationships between the
fluxes versus the optical depth. Therefore, the observational results
are in good accord with our results. However, more quantitative comparisons
would require more realistic models using external radiation fields.

If observations are made over a wide range of Galactic coordinates,
as in the observations of the diffuse Galactic light (DGL), rather
than a well-defined single cloud, the obtained correlation would be
a result of the complex superposition of many clouds with various
values of $\tau_{{\rm H}}$. To understand the general properties
of such observations, we assume a virtual system consisting of clouds
with the 10 different $\tau_{{\rm H}}$ values considered in Figure
\ref{correlation}. We also assume that the clouds are exposed to
radiation sources with the same luminosity.

\begin{figure}[t]
\begin{centering}
\includegraphics[scale=0.68]{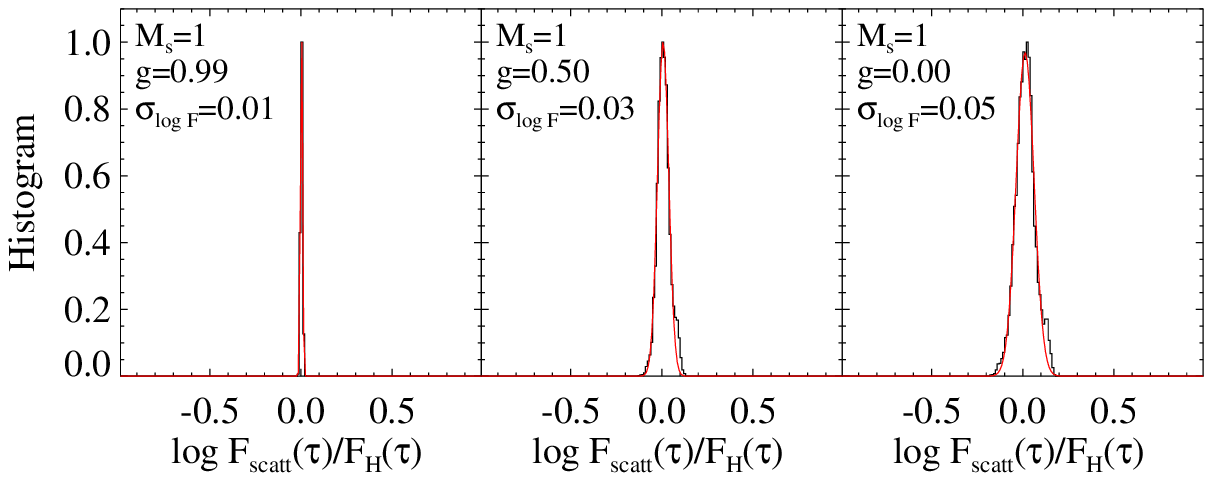} 
\par\end{centering}

\begin{centering}
\medskip{}

\par\end{centering}

\begin{centering}
\includegraphics[scale=0.68]{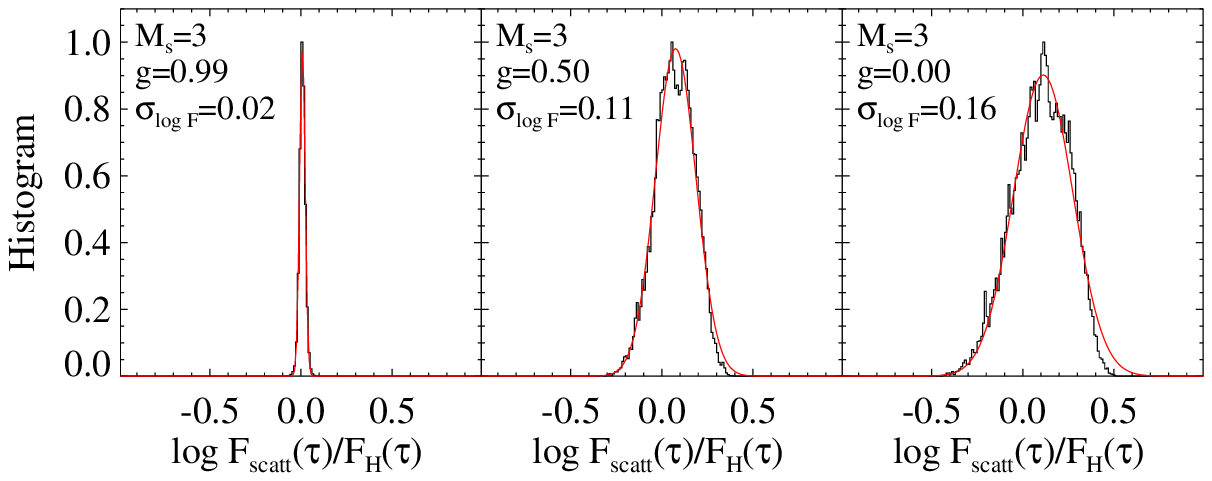} 
\par\end{centering}

\caption{\label{lognormal}Histograms of the deviations (spreads) of the scattered
fluxes ($F_{{\rm scatt}}$) from the results of the uniform density
clouds ($F_{{\rm H}}$). The red curves denote the best-fit lognormal
functions. Standard deviations ($\sigma_{\log F}$) of the lognormal
functions are also shown.}
\end{figure}

In this context, Figure \ref{correlation} can be regarded as the
observed correlation in the virtual system. We can find, in Figure
\ref{correlation}, a general correlation between the scattered light
$F_{{\rm scatt}}$ and the optical depth $\tau$ for a wide range
of optical depths. This indicates that the mean value of the scattered
flux $\left\langle F_{{\rm scatt}}\right\rangle $ is roughly proportional
to $\tau$. However, large scatter in the correlation is also found.
We also note that the spread, measured in terms of the standard deviation
$\sigma_{F}$, of $F_{{\rm scatt}}$ increases with $\tau$. In other
words, the spread $\sigma_{F}$ is proportional to $\tau$ and is
thus proportional to the mean value of the scattered flux $\left\langle F_{{\rm scatt}}\right\rangle $.
The increase of the flux spread with the mean flux is a property of
a lognormal function. We therefore obtained histograms of the scattered
fluxes ($F_{{\rm scatt}}$) normalized to the fluxes ($F_{{\rm H}}$)
of uniform density models, which can be regarded as average fluxes.
The resulting PDFs are lognormal, as shown in Figure \ref{lognormal}.
In the figure, the best-fit lognormal functions and the standard deviations
($\sigma_{\ln F}$) of the logarithmic fluxes are also shown. The
standard deviation increases with an increase in $M_{{\rm s}}$ and
a decrease in $g$. The same trend is also apparent in Figure \ref{correlation}.

\citet{Seon2013} presented the standard deviations of the dust column
density as measured from molecular clouds and the FUV background intensities.
The standard deviations $\sigma_{{\rm \log}F}$ for $M_{{\rm s}}=3$
in Figure \ref{lognormal} are well within the range found from molecular
clouds but are smaller than those estimated from the FUV background.
The smaller standard deviations in our models compared to the FUV
observations may be due to the lack of consideration of realistic
radiation fields in our models or may imply that the dust clouds responsible
for the FUV background are more turbulent than our cloud models. However,
we note that our models are too simplified to be quantitatively compared
with the observed results of the DGL.

Some basic properties similar to our results were found in previous
investigations using a rather simplified medium. \citet{Witt1996}
dealt with the escape of stellar radiation in a clumpy, two-phase
medium. They found that the sharp central peak of the radial intensity
distribution observed in the uniform density cloud disappeared in
the clumpy medium. \citet{Mathis2002} used hierarchically clumped
clouds and found that the optical properties of grains are poorly
constrained by observations of reflection nebulae, mainly due to a
very large spread in both the scattered fluxes and stellar extinctions.
Our results are qualitatively consistent with their results. However,
our models show less severe departures from uniform density models.
In \citet{Mathis2002}, the scattered fluxes at the sightlines with
the average optical depths ($\tau_{{\rm ext}}=2$ in their Figures
1 and 2) were always lower than those of uniform density models. However,
in our results, the scattered fluxes at $\tau=\tau_{{\rm H}}$ are
always centered around the values of uniform density models. The density
structures in \citet{Mathis2002} included a large number of empty
cells. Their models in which a constant density was added to all cells
to avoid zero density gave rise to higher scattered fluxes than the
models with empty cells, mainly due to the contribution of scattered
fluxes from the constant density cells. A lognormal density PDF in
our models yielded even higher scattered fluxes than the clouds with
the constant density cells in \citet{Mathis2002}. We also obtained
no significant difference, unlike their results, when the central
source was embedded within a relatively dense region. This is because
we mainly considered optically thin clouds, while \citet{Mathis2002}
covered larger optical depths.

\section{SUMMARY}

Dust scattering in a turbulent, dusty ISM yielded large scatter in
the correlation plots between the scattered flux and the optical depth.
The slopes of the correlation plots from turbulent media are in general
shallower than those expected from uniform density models. The effect
becomes more significant as $M_{{\rm s}}$ increases and $g$ decreases.
The correlation between the scattered fluxes at two different wavelengths
is found to be tighter than the plots of the scattered fluxes versus
the optical depth. These results are in good accord with the observed
results of the scattered light in the optical and FUV wavelengths.
Given that micron-size dust grains are rare in the diffuse ISM, the
scattering in NIR wavelengths (especially at $\lambda\gtrsim2\mu{\rm m}$)
would be essentially isotropic, as shown in \citet{Draine03}. Therefore,
the scattered light in the NIR would show a much weaker correlation
with optical depth than the observations at optical and UV wavelengths.
Our future works will deal with various configurations between clouds
and radiation field, including the case of a cloud exposed to an external
radiation field.

\pagebreak{}

\begin{thebibliography}{Kritsuk, Norman, \& Padoan(2006)}
\bibitem[Berkhuijsen \& Fletcher(2008)]{Berkhuijsen2008}Berkhuijsen,
E. M., \& Fletcher, A. 2008, \mnras, 390, L19

\bibitem[Bianchi et al.(2000)]{Bianchi2000}Bianchi, S., Ferrara,
A., Davies, J. I., \& Alton, P. B. 2000, \mnras, 311, 601

\bibitem[Bianchi et al.(1996)]{Bianchi1996}Bianchi, S., Ferrara,
A., Giovanardi, C. 1996, \apj, 465, 127

\bibitem[Bowyer(1991)]{Bowyer91}Bowyer, S. 1991, \araa, 29, 59

\bibitem[Brandt \& Draine(2012)]{Brandt2012}Brandt, T. D., \& Draine,
B. T. 2012, \apj, 744, 129

\bibitem[Burkhart \& Lazarian(2012)]{Burkhart2012}Burkhart, B., \&
Lazarian, A. 2012, \apjl, 755, L19

\bibitem[Cashwell \& Everett(1959)]{Cashwell1959}Cashwell, E. D.,
\& Everett, C. J. 1959, A Practical Manual on the Monte Carlo Method
for Random Walk Problems (New York : Pergamon)

\bibitem[Choi et al.(2013)]{Choi2013}Choi, Y.-J., Min, K.-W., Seon,
K.-I., Lim, T.-H., Jo, Y.-S., \& Park, J.-W. \apj, 2013, 774, 34

\bibitem[De Looze et al.(2012)]{DeLooze_Somberero2012}De Looze, I.,
Baes, M., Fritz, J., \& Verstappen, J. 2012, \mnras, 419, 895

\bibitem[Draine(2003)]{Draine03}Draine, B. T. 2003, \apj, 598, 1017

\bibitem[Elmegreen(1997)]{Elmegreen1997}Elmegreen, B. G. 1997, \apj,
477, 196

\bibitem[Elmegreen(2002)]{Elmegreen2002}Elmegreen, B. G. 2002, \apj,
564, 773

\bibitem[Elmegreen \& Scalo(2004)]{Elmegreen2004}Elmegreen, B. G.,
\& Scalo, J. 2004, \araa, 42, 211

\bibitem[Federrath et al.(2008)]{Federrath2008}Federrath, C., Klessen,
R. S., \& Schmidt, W. 2008, \apjl, 688, L79

\bibitem[Federrath et al.(2010)]{Federrath2010}Federrath, C., Roman-Duval,
J., Klessen, R. S., Schmidt, W., \& Mac Low, M.-M., 2010, \aap, 512,
A81

\bibitem[Froebrich \& Rowles(2010)]{Froebrich2010}Froebrich, D.,
\& Rowles, J. 2010, \mnras, 406, 1350

\bibitem[G{\'o}rski et al.(2005)]{Gorski05}G{\'o}rski, K. M., Hivon,
E., Banday, A. J., et al. 2005, \apj, 622, 759

\bibitem[Guhathakurta \& Tyson(1989)]{Guhathakurta1989}Guhathakurta,
P., \& Tyson, J. A. 1989, ApJ, 346, 773

\bibitem[Ienaka et al.(2013)]{Ienaka2013}Ienaka, N., Kawara, K.,
Matsuoka, Y., et al. 2013, \apj, 767, 80

\bibitem[Jo et al.(2012)]{Jo2012}Jo, Y.-S., Min, K. W., Lim, T.-H.,
\& Seon, K.-I. 2012, \apj, 756, 38

\bibitem[Klessen(2000)]{Klessen2000}Klessen, R. S. 2000, \apj, 535,
869

\bibitem[Kim \& Ryu(2005)]{Kim05}Kim, J., \& Ryu, D. 2005, \apjl,
630, L45

\bibitem[Kritsuk, Norman, \& Padoan(2006)]{Kritsuk06}Kritsuk, A.
G., Norman, M. L., \& Padoan, P. 2006, \apjl, 638, L25

\bibitem[Lim et al.(2013)]{Lim2013}Lim, T.-H., Min, K.-W., \& Seon,
K.-I. 2013, \apj, 765, 107

\bibitem[Lombardi et al.(2008)]{Lombardi2008}Lombardi, M., Alves,
J., \& Lada, C. J. 2008, \aap, 489, 143

\bibitem[Malinen et al.(2013)]{Malinen2013}Malinen, J., Juvela, M.,
Pelkonen, V. M., \& Rawlings, M. G. 2013, \aap, 558, A44

\bibitem[Mathis et al.(1983)]{MMP1983}Mathis, J. S., Mezger, P. G.,
\& Panagia, N. 1983, \aap, 128, 212.

\bibitem[Mathis et al.(2002)]{Mathis2002}Mathis, J. S., Whitney,
B. A., \& Wood, K. 2002, \apj, 574, 812

\bibitem[Ostriker et al.(2001)]{Ostriker01}Ostriker, E. C., Stone,
J. M., \& Gammie, C. F. 2001, \apj, 546, 980

\bibitem[Padoan et al.(1997)]{Padoan97}Padoan, P., Jones, B. J. T.,
\& Nordlund, \AA. 1997, \apj, 474, 730

\bibitem[Padoan et al.(2004)]{Padoan04}Padoan, P., Jimenez, R., Juvela,
M., \& Nordlund, \AA. 2004, \apjl, 604, L49

\bibitem[Schiminovich et al.(2001)]{Schiminovich2001}Schiminovich,
D., Friedman, P. G., Martin, C., \& Morrissey, P. F. 2001, \apjl,
563, L161

\bibitem[Seon(2009)]{Seon2009}Seon, K.-I. 2009, \apj, 703, 1159

\bibitem[Seon(2012)]{Seon2012}Seon, K.-I. 2012, \apjl, 761, L17

\bibitem[Seon(2013)]{Seon2013}Seon, K.-I. 2013, \apj, 772, 57

\bibitem[Seon et al.(2011)]{Seon2011a}Seon, K.-I., Edelstein, J.,
Korpela, E. J., et al. 2011, \apjs, 196, 15

\bibitem[Seon \& Witt(2012)]{SeonWitt2012}Seon, K.-I., Witt, A. N.
2012, \apj, 758, 109

\bibitem[Stalevski et al.(2012)]{Stalevski2012}Stalevski, M., Fritz,
J., Baes, M., Nakos, T., \& Popovi{\'c}, L. \v{C}. 2012, \mnras,
420, 2756

\bibitem[Stutzki et al.(1998)]{Stutzki1998}Stutzki, J., Bensch, F.,
Heithausen, A., Ossenkopf, V., \& Zielinsky, M. 1998, \aap, 336,
697

\bibitem[V{\'a}zquez-Semadeni(1994)]{Vazquez94}V\'azquez-Semadeni,
E. 1994, \apj, 423, 681

\bibitem[Voss(1988)]{Voss1988}Voss, R. 1988, in The Science of Fractal
Images, ed. H. O. Peitgen, \& D. Saupe (New York : Springer), 22

\bibitem[Witt(1977)]{Witt1977I}Witt, A. N. 1977, \apjs, 35, 1

\bibitem[Witt et al.(1982)]{Witt1982}Witt, A. N., Walker, G. A. H.,
Bohlin, R. C., \& Stecher, T. P. 1982, \apj, 261, 492

\bibitem[Witt et al.(1992)]{Witt1992}Witt, A. N., Thronson, H. A.,
\& Capuano, J. M. 1992, \apj, 393, 611

\bibitem[Witt \& Gordon(1996)]{Witt1996}Witt, A. N., \& Gordon, K.
D. 1996, \apj, 463, 681

\bibitem[Witt et al.(1997)]{Witt1997}Witt, A. N., Friedmann, B. C.,
\& Sasseen, T. P. 1997, \apj, 481, 809

\bibitem[Witt \& Gordon(2000)]{Witt2000}Witt, A. N., \& Gordon, K.
D. 2000, \apj, 528, 799

\bibitem[Witt et al.(2008)]{Witt2008}Witt, A. N., Mandel, S., Sell,
P. H., Dixon, T., \& Vijh, U. P. 2008, \apj, 679, 497

\bibitem[Yusef-Zadeh et al.(1984)]{Yusef-Zadeh1984}Yusef-Zadeh, F.,
Morris, M., \& White, R. L. 1984, \apj, 278, 186

\end{thebibliography}
\end{document}